\documentclass[showpacs,preprintnumbers,amsmath,amssymb]{revtex4}

\usepackage[latin1]{inputenc}
\usepackage{makeidx}
\usepackage{layout}        
\usepackage{color}
\usepackage{graphicx}
\usepackage{dcolumn}
\usepackage{bm}

\newcommand{\btr}{\blacktriangleright}
\newcommand{\btl}{\blacktriangleleft}

\begin{document}
\preprint{preprint}

\title{Diffusion rates of Cu adatoms on Cu(111) in the presence of an
adisland nucleated at FCC or HCP sites}

\author{     Mihai-Cosmin Marinica$^*$, Cyrille Barreteau$^*$, Daniel Spanjaard$^{\dag}$ and Marie-Catherine Desjonqu\`eres$^*$ }
\affiliation{$^*$CEA Saclay, DSM/DRECAM/SPCSI, B\^atiment 462, F-91191 Gif sur Yvette, France }

\affiliation{$^{\dag}$Laboratoire de Physique des Solides,
             Universit\'e Paris Sud, Batiment 510, F-91405 Orsay, France}

\date{\today}
\begin{abstract}

The surface diffusion of Cu adatoms in the presence of an adisland
at FCC or HCP sites on Cu(111) is studied using the EAM potential
derived by Mishin {\it et al.} [Phys. Rev. B {\bf 63} 224106
(2001)]. The diffusion rates along straight (with close-packed
edges) steps with (100) and (111)-type microfacets (resp. step A
and step B) are first investigated using the
transition state theory in the harmonic approximation. It is found
that the classical limit beyond which the diffusion rates follow
an Arrhenius law is reached above the Debye temperature. The
Vineyard attempt frequencies and the (static) energy barriers are
reported. Then a comparison is made with the results of more
realistic classical molecular dynamic simulations which also
exhibit an Arrhenius-like behavior. It is concluded that the
corresponding energy barriers are completely consistent with the
static ones within the statistical errors and that the diffusion
barrier along step B is significantly larger than along step A. In
contrast the prefactors are very different from the Vineyard
frequencies. They increase with the static energy barrier in
agreement with the Meyer-Neldel compensation rule and this increase is well
approximated by the law proposed by Boisvert {\it et al.} [Phys.
Rev. Lett. {\bf 75} 469 (1995)]. As a consequence, the remaining
part of this work is devoted to the determination of static energy
barriers for a large number of diffusion events that can occur in
the presence of an adisland. In particular, it is found that the
corner crossing diffusion process for triangular adislands is
markedly different for the two types of borders (A or B). From
this set of results the diffusion rates of the most important
atomic displacements can be predicted and used as input in Kinetic
Monte-Carlo simulations.

\end{abstract}

\maketitle

\section{Introduction}
The diffusion of adatoms on metal surfaces is still the subject of
very active research \cite{Zhang97,Giesen01}. Indeed, it plays a crucial role in crystal
growth which is important to master in view of the applications,
for instance in nanotechnologies. The more the growth is
understood at an atomic level, the more the fabrication processes
can be controlled in order to obtain a better quality of the
device performances.

As a starting point, homo-epitaxial systems can be used as models
since effects such as lattice mismatch or inter-diffusion
processes are excluded. A large number of theoretical and
experimental works has been devoted to homo-epitaxial growth,
e.g., on Pt \cite{Liu93,Villarba94,Jacobsen96,Michely93,Golzhauser96},
 Cu \cite{Wulfhekel96,Schlosser00,Giesen03}, Ag \cite{Yu96, Fichthorn00},
 Rh \cite{Papadia96,Wang90,Maca00}, Al \cite{Stumpf96,Ruggerone97,Ovesson99}...
 In this respect, the understanding of
adisland shapes is of fundamental importance since they can
indirectly influence the growth mode. The (111) surface of FCC
metals is particularly interesting due to the observation of
fractal as well as two (2D) or three (3D) dimensional compact
adislands depending on the temperature
\cite{Michely93,Jacobsen96,Li96}. Actually, the adisland
morphology and its evolution with temperature result from a
competition between thermodynamics and kinetic phenomena depending
on the activation (or not) of various diffusion processes. The
shape of adislands is governed by thermodynamics when the
temperature is such that the adisland is able to relax to its
equilibrium configuration during the time interval separating the
incorporation of two consecutive atoms. In this case the shape of
adislands on a (111) FCC surface is determined by the ratio of two
step energies since two types of steps with close-packed edges
exist on this surface: the A step with a ledge of (100)-type
orientation and B step with a ledge of (111)-type. The anisotropy
ratio of the two step energies
$E_{\mbox{step}}^A/E_{\mbox{step}}^B$ is most often very close to
unity (for instance, using the Mishin {\it et al.}\cite{Mishin01}
potential for Cu, we have found $E_{\mbox{step}}^A$=263meV and
$E_{\mbox{step}}^B$=265meV \cite{Marinica04}) and therefore the
adisland shape derived from the Wulff theorem is an almost regular
hexagon exhibiting three-fold symmetry with edges corresponding to
the smallest step energy slightly longer than the other ones
\cite{Raouafi02}. When kinetic effects dominate, the adisland
morphology depends on the flux of atoms impinging on the surface
as well as on the rates of adatom diffusion on terraces, along
steps, around adisland corners and of adatom attachment to
(detachment from) the adisland.

The most efficient tool to simulate the evolution of the adisland
morphology is the Kinetic Monte-Carlo 
(KMC \cite{BKL75,Jensen99,Pomeroy02}) method which needs as
input the various rates of all the elementary atomic processes.
Two types of processes are essential in this respect: the 
diffusion along steps (and its anisotropy between step A and step B), and the
corner crossings to go from one edge to another. At low
temperature when none of these processes are activated one expects
a fractal growth if the impinging flux is not too small since
diffusion is rapid on a flat (111) FCC surface (e.g. the
activation energy is only 40meV for Cu). When these two processes
occur, compact adislands can grow. An intermediate situation is
also possible corresponding to an activation of step diffusion
while corner crossing is still frozen. However other physical
properties may have some importance: for instance, the
step on which adatoms or dimers preferentially bind and the kink
dissociation process play a role on the adisland shape which can
vary with temperature from triangles limited by A steps, to
triangles limited by B steps with intermediate hexagon-like
patterns \cite{Liu93,Jacobsen96}. The various possible sequences are
extremely dependent on the details of the energy profile and the
modification of a single barrier can lead to very different
adisland shapes. Moreover the diffusion across a step plays a
crucial role in the crystal growth mode (Volmer-Weber, Stransky-Krastanov or
Frank Van der Merwe). Indeed during growth adatoms
can be deposited either on a wide terrace or on a preformed
adisland. In the former case, the atom will diffuse until it
sticks to the adisland, whereas in the latter case the adatom
needs very often to overcome an extra activation energy (Schwoebel
barrier) to incorporate to the descending step edge. Depending on
temperature a 3D growth can thus be initiated. Therefore we have
carried out a very detailed analysis of the various possible
diffusion processes of a copper adatom in the vicinity of an
adisland. Two types of epitaxy will be considered in which adatoms
occupy either normal FCC sites or faulted HCP sites. Actually the
stacking fault energy is very small in copper and adisland
nucleation has been observed at FCC sites on the nominally flat
surface and at HCP sites on the Cu(21,21,23) vicinal
surface\cite{Giesen03}.

Many experiments based on field ion\cite{Stolt76,Antczak04} and
scanning tunnelling \cite{Linderoth97,Giesen01,Repp03}
microscopies have been devoted to the determination of surface
diffusion coefficients. However their interpretation may be
somewhat tricky since, apart from peculiar cases, several
elementary processes are involved. On the theoretical side,
diffusion rates have been deduced either from Transition State
Theory (TST) in the harmonic approximation
\cite{Kurpick97a,Kurpick97b,Kurpick98,Marinica04} (TST-HA), or using
classical thermodynamical integration\cite{Boisvert98} (TI) (see Appendix A) or else
from classical molecular dynamic (MD)
simulations\cite{Ferrando94,Boisvert95,Boisvert96,Boisvert97,Montalenti99}. In the
classical limit all theoretical approaches conclude that an
Arrhenius law, $\Gamma=\Gamma_0\exp(-\Delta E/k_BT)$, fits accurately
the evolution of the diffusion rates with temperature
in good agreement with experiments. On the basis of MD simulations
it has been suggested that the barrier $\Delta E$ for diffusion
along steps on Ag(111) and Au(111) might be different from the
static barrier \cite{Ferrando94} whereas Boisvert {\it et al.}
\cite{Boisvert98} found that the value of the static barrier for
Cu on Cu(100) lies inside the error bars of the results obtained
both with MD and TI methods. Actually a very good accuracy on
$\Delta E$ (better than a few $10^{-2}$eV) would need huge
simulation times. Furthermore the Arrhenius law being obviously
approximate, the separation between a prefactor and an exponential
term is somewhat arbitrary as will be discussed in Sec.IIB.

From the above remarks, it is clear that any investigation of
diffusion processes should begin with the determination of the
(static) potential energy barriers. A limited number of barriers
concerning high symmetry systems have been calculated using
ab-initio codes\cite{Ratsch98} but these codes are too computer time demanding
when the symmetry is low, i.e., when the number of elementary
diffusion processes is large. One must then rely on semi-empirical
potentials such as those derived from Effective Medium Theory 
(EMT)\cite{Stoltze94}, Embedded Atom Model (EAM)\cite{Karimi95,Trushin97} or
Second Moment Approximation (SMA)\cite{Evangelakis98}. Even with these simple models,
unknown barriers are sometimes deduced using some approximations
\cite{Schlosser00}. Even though these approximations are
reasonable some effects can be missed as we will see in the
following.

The aim of this work is the determination of the diffusion rates
corresponding to most of the elementary diffusion processes
that may occur during the growth of a single 2D adisland on
Cu(111). After having briefly presented the potential
used (Sec.IIA) we study  the diffusion of an adatom along step
A and step B (Sec.III) using the two methods described in Sec.IIB, namely
the TST-HA approach (in the classical limit) and MD simulations.
We show that the static activation barrier accounts quite well
for the results of MD simulations. However, the prefactors
obtained in the latter method are different from those
given by TST-HA. Furthermore they are quite consistent with
the Meyer-Neldel law proposed by Boisvert {\it et al.}
\cite{Boisvert95}. As a consequence, in the remaining part
of the paper, we limit ourselves to the determination of
the static barriers for other diffusion events in the
presence of straight steps (Sec.IV) or in the vicinity of
steps with defects, for instance around corners and kinks
(Sec.V). Conclusions are drawn in Sec.VI. Finally Appendix A
briefly summarizes the TST-HA theory and the TI method.


\section{Formalism}

\subsection{The potential}

We have used in this work the EAM potential derived by Mishin {\it
et al.}\cite{Mishin01} for copper. In this model the total energy
of an assembly of N atoms with respect to that of N isolated atoms
is written as a function of all interatomic distances $r_{ij}$:

\begin{equation}
E=\frac{1}{2}\sum_{i,j=1}^N V(r_{ij})+\sum_{i=1}^N F(\rho_i)
\label{etot}
\end{equation}

\noindent where $\rho_i$ can be interpreted as a function
proportional to the electron density induced at site $i$ by the
neighbors, i.e.:

\begin{equation}
\rho_i=\sum_{j\neq i}\rho(r_{ij}). \label{rho}
\end{equation}

\noindent The proportionality factor is chosen such that
$\rho_i=1$ for a bulk atom at equilibrium. The chosen reference energy implies
that $V(r)$ and $\rho(r)$ vanish when $r$ tends to infinity and
$F(0)=0$. The functions $V(r)$, $\rho(r)$, and $F(\rho)$ are
fitted parameterized functions which, on the whole, contain 26
parameters. The 26 parameters are required to give exactly the
bulk equilibrium lattice parameter, cohesive energy and bulk
modulus and to fit selected properties of copper taken from
experiments or obtained by {\it ab-initio} calculations. These
properties are mainly related to the bulk phase and include in
particular the stacking fault energy which is of a peculiar
importance in the present work. More details can be found in
Ref.~\cite{Mishin01}. In addition, Mishin {\it et al.} have shown
that their potential was able to reproduce many physical
quantities not included in the fitting data base and, in
particular, the surface energies of low index surfaces. Moreover
we have already used this potential in a previous work \cite{Marinica04}. First we
verified that it gives other surface properties, like the step and
kink energies, with numerical values very close to experimental
data. Then we used it to study the diffusion of monomers, dimers
and trimers on Cu(111) and our results were in very good agreement
with STM observations. We are thus very confident in using this
potential for the present problem.

\subsection{Determination of diffusion coefficients along straight steps}

The diffusion coefficients along straight steps (i.e., with a
close-packed edge) have been calculated using two techniques. On
the one hand, they can be derived from the TST\cite{Wahnstrom90}
by combining the determination of the minimum energy path
and the calculation of the vibrational free energy in the
framework of the HA and, on the other
hand, using MD simulations. Indeed, as will
be seen in the following, the most frequent diffusion event for an
adatom attached to a straight step corresponds to jumps along this
step since the barriers corresponding to other events are much
higher. Thus this diffusion is essentially a one-dimensional
problem.

\subsubsection{Diffusion coefficients of an adatom along straight steps from TST-HA}

From TST the surface diffusion coefficient $D(T)$ along straight
steps at temperature T is related to the diffusion frequency (or
diffusion rate) $\Gamma_{TST}(T)$ by:

\begin{equation}
D(T)=\frac{1}{2}a^2\Gamma_{TST}(T)
\label{DTTST}
\end{equation}

\noindent when the diffusion is assumed to proceed by uncorrelated
jumps of length $a$ between neighboring adsorption sites along the step
(note that $a$ increases slowly with T when thermal expansion is taken
into account but in practice this effect is quite negligible). The
diffusion frequency $\Gamma_{TST}(T)$ is given by:

\begin{equation}
\Gamma_{TST}(T)=2\frac{k_BT}{h}\exp(-\Delta F/k_BT)
\label{GTST}
\end{equation}

\noindent where $\Delta F$ is the difference of total free energy
between the saddle point and the starting adsorption configuration
(stable or metastable equilibrium). The factor 2 arises from the
number ($n_c=2$) of diffusion channels, i.e., $\Gamma_{TST}(T)$ is
the diffusion frequency in {\it both} directions along the step.
$\Delta F$ can be split into two contributions:

\begin{equation}
\Delta F=\Delta E+\Delta F_{vib} \label{DF}
\end{equation}

\noindent $\Delta E$ is the potential energy barrier, i.e., the
difference in total energy (Eq.\ref{etot}) between the saddle
point and the equilibrium configuration. This static diffusion
barrier $\Delta E$ can be calculated by determining the minimum
energy path using the Ulitsky-Elber algorithm \cite{Ulitsky90}
which we already used with success in our study of the surface
diffusion on the flat Cu(111) surface \cite{Marinica04}. For
technical details the reader is referred to this latter reference.
The quantity $\Delta F_{vib}=\Delta U_{vib}-T\Delta S_{vib}$ is
the contribution of vibrations to the variation of the free energy
(with obvious notations). In the HA (see Appendix), $\Delta F_{vib}$ is given by:

\begin{equation}
\Delta
F_{vib}=k_BT\int_0^{\nu_{max}}\ln(2\sinh(\frac{h\nu}{2k_BT}))\Delta
n(\nu)d\nu \label{DFvib}
\end{equation}

\noindent where $\Delta n(\nu)$ is the difference between the
vibrational densities at the saddle point and at equilibrium which
are computed from the eigenvalues of the dynamical matrix.
 Note that at the saddle point one of the eigenvalues $\nu_p^2$
is negative. This eigenvalue is excluded from the vibration
density and, consequently:

\begin{equation}
\int_0^{\nu_{max}}\Delta n(\nu)d\nu=-1.
\end{equation}

\noindent Thus in this TST-HA model, the diffusion rate is:

\begin{equation}
\Gamma_{TST}^{HA}(T)=4\frac{k_BT}{h} \frac{\displaystyle
\prod_{p=1}^{3N}\sinh(\frac{h\nu_p^e}{2k_BT})}
     {\displaystyle \prod_{p=1}^{3N-1}\sinh(\frac{h\nu_p^s}{2k_BT})}\exp(-\Delta E/k_BT)
      =\Gamma_0^{HA}(T)\exp(-\Delta E/k_BT)
\label{Dsh}
\end{equation}

\noindent if we call $\nu_p^e$ and $\nu_p^s$ the (real)
eigenfrequencies for the equilibrium and saddle point
configurations and $N+1$ the total number of atoms, i.e.,
including the adatom. Note that the three free translational modes
(of zero frequency) have been excluded from the spectra. It is
seen that the prefactor $\Gamma_0^{HA}(T)$ depends in general on T
and, consequently, $\Gamma_{TST}^{HA}(T)$ does not follow an
Arrhenius law.

Let us examine the limits of low and high temperatures. In the low
temperature limit $\Delta S_{vib}\rightarrow 0$ and $\Delta
U_{vib}(T)=\Delta U_{vib}(0)+\mathcal{O}(T^4)$ ~\cite{Pines64} so
that:

\begin{equation}
\Gamma_{TST}^{HA}(T) \simeq
2\frac{k_BT}{h}\exp(-(\Delta E+\Delta
U_{vib}(0))/k_BT) \label{DT0}
\end{equation}

\noindent
In the high temperature limit we get:

\begin{equation}
\Gamma_{TST}^{HA}(T)=2\nu_0~\exp~(-\Delta E/k_BT) \label{DTinf}
\end{equation}

\noindent in which $\nu_0$ is the Vineyard attempt frequency given by \cite{Vineyard57}:

\begin{equation}
\nu_0=\frac{\displaystyle  \prod_{p=1}^{3N}\nu_p^e}{\displaystyle \prod_{p=1}^{3N-1}\nu_p^s}.
\label{nu0}
\end{equation}

\noindent Thus, in this limit, $\Gamma_{TST}^{HA}(T)$ follows an Arrhenius law with a diffusion
barrier $\Delta E$ and a prefactor $2\nu_0$.

In this TST-HA approach the deviation from the Arrhenius law comes
from quantum effects. We will see in the following that these
effects become negligible above the Debye temperature ($T_D=343K$
for Cu \cite{Kittel71}). Accordingly the classical molecular
dynamics  approach, which we will briefly describe in the next
subsection, is fully justified above $T_D$.

\subsubsection{Diffusion coefficients along straight steps from MD
simulations}

The motion of the N+1 atoms is studied by MD simulations in which
the Newton equations of motion in the potential given by
Eq.~\ref{etot} is solved using the Verlet algorithm in its
velocity form. The simulation is carried out in the temperature
range 350K-600K and no correction of the temperature is needed
since quantum effects can be neglected ~\cite{Wang90b}. The
thermal expansion of the Cu crystal is derived from MD simulations in the
bulk and the nearest neighbor distance is fitted by   
$a=2.553+2.516 10^{-5}T+1.1939 10^{-8}T^2$ in \AA. After an
equilibration period, the motion of the adatom is observed during
another period $t_{run}$ and the trajectory of the adatom is
recorded. This trajectory is analyzed by assuming that the
diffusion proceeds by uncorrelated and discrete jumps. At this
point, we must emphasize the differences between MD and TST. The
basic assumption in TST is that each crossing of a dividing
surface containing the saddle point corresponds to a diffusion
event and thus ``recrossing effects'' are ignored~\cite{Wahnstrom90}. This is not the
case in MD since recrossing events can be identified, i.e., they
occur when the adatom turns back in a very short time after having
crossed the dividing surface. Moreover in Eq.\ref{DTTST} each
diffusion event is assumed to be a jump of length $a$ while long
jumps of length $na$ can occur in MD. Finally, anharmonic effects
are included since the forces are calculated from the exact
expression of the potential.

The diffusion coefficient is given by the Einstein formula:

\begin{equation}
D(T) = \lim_{t \rightarrow \infty}\frac{<({\bf r}(t))^2>}{2t}
\label{Dmd}
\end{equation}

\noindent where

\begin{equation}
{\bf r}(t)=\sum_{i=1}^{N_S} n_i{\bf r}_i,
\end{equation}

\noindent $N_S$ is the total number of jumps during the time
interval $t$, $n_i{\bf r}_i$ is the displacement vector
corresponding to the $i^{th}$ jump ($n_i=1,2,3...$ for a simple,
double, triple... jump, respectively, and $r_i=a$) and, for
uncorrelated jumps (i.e., $<{\bf r}_i.{\bf r}_j>=0$), if $t_{run}$
is large enough:

\begin{equation}
D(T)=\frac{a^2}{2}\sum_{i=1}^{N_S}\frac{n_i^2}{t_{run}}=\frac{a^2}{2}\sum_n
\frac{N_n}{t_{run}}n^2=\frac{a^2}{2}\Gamma^{MD}(T)
\label{Djump}
\end{equation}

\noindent where $N_n$ is the number of jumps of length $na$
observed during the period $t_{run}$.

\begin{figure}[!fht]
\begin{center}
\includegraphics*[scale=0.5,angle=0]{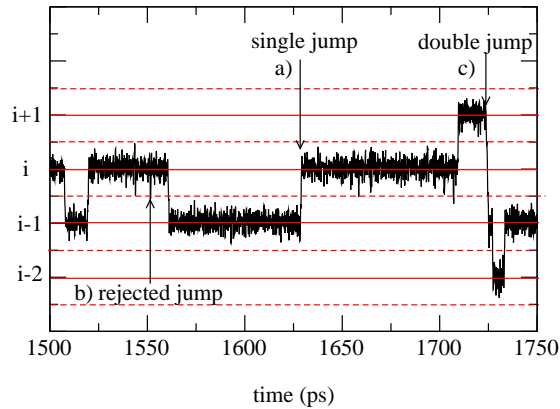}
\end{center}
\caption{Typical time evolution of the position of an adatom along
a step. The adatom starts from site $i$ and the unit along the
ordinate axis is the nearest neighbor spacing $a$. The adatom is
considered to be at site $i$ if its coordinate lies in the interval
$[i-1/2,i+1/2]$. The rejected,
single and double jumps are marked by arrows. }
\label{fig:thermalization}
\end{figure}

The discrimination between the accepted and rejected diffusion
events and between the different jump lengths is done with the
same criterion as in the work of Ferrando and Tr\'eglia
\cite{Ferrando94}. Let us call $\tau_{th}$ the thermal time given
by:

\begin{equation}
\tau_{th}=a(\frac{m}{k_BT})^{1/2}
\end{equation}

\noindent where $m$ is the mass of the adatom, i.e., $\tau_{th}$
is simply the time to go over the lattice spacing for a particle
with an energy $k_BT$, for example $\tau_{th}=1ps$ at T=500K. The
diffusing atom is considered to have resided at a given site if
the time spent on this site is longer than $\tau_{th}$. This is
illustrated in Fig.\ref{fig:thermalization}. For instance the case
$(a)$ corresponds to a single jump since the time spent at sites
$i-1$ and $i$ is longer than $\tau_{th}$, whereas no jump is
recorded in the case $(b)$ since the time spent at $i+1$ is too
short (recrossing event). Finally the case $(c)$ corresponds to a
double jump since the time spent at site $i$ is too short.

The usual method to analyze the temperature dependence of
$\Gamma^{MD}$ is to draw an Arrhenius plot, i.e.,
$\ln\Gamma^{MD}(T)$ vs $1/k_BT$
\cite{Ferrando94,Boisvert95,Boisvert96,Boisvert97,Boisvert98,Montalenti99}.
Most often a quasi-linear behavior is observed, the larger the
number of recorded diffusion events (i.e., the smaller the
statistical error), the smaller the deviation from linearity. As a
consequence, $\Gamma^{MD}(T)$ can be approached by an Arrhenius
law:

\begin{equation}
\Gamma^{MD}(T)=\Gamma^{MD}_0~\exp~(-\Delta E^{MD}/k_BT)
\label{arrh}
\end{equation}

\noindent in which the parameters $\Gamma^{MD}_0$ and $\Delta
E^{MD}$ are obtained by a least mean square fit.

It is then found that $\Delta E^{MD}$ differs slightly from the
static barrier $\Delta E$. This small discrepancy may be due to
statistical errors. Actually, as we will illustrate in the
following (see Fig.\ref{fig:dynamic_hopping_rate_AB_fcchcp}),
if the fit is carried out by setting $\Delta
E^{MD}=\Delta E$ the corresponding straight line is also contained
inside the error bars. Furthermore, if real physical effects were
responsible for the variation between $\Delta E^{MD}$ and $\Delta
E$, then ($\Delta E^{MD}-\Delta E$) should be a function of
temperature. Consequently the Arrhenius law is not strictly obeyed and
the splitting of $\Gamma^{MD}$ into a prefactor and an exponential
is somewhat arbitrary. However if the variation of ($\Delta E^{MD}-\Delta E$)
is of the first order in T, then the Arrhenius law remains almost strictly obeyed
with a slope equal to $\Delta E$.

Actually, using TI (see Appendix) Boisvert
et al \cite{Boisvert98} have shown that the diffusion frequency
can be written as an Arrhenius-type law in which the prefactor
$\Gamma_0^{TI}$ and the barrier $\Delta E^{TI}$ are almost
independent of temperature in the range $100-800K$ for surface
diffusion of Cu on Cu(100). Moreover, by comparing the TI
method with MD simulations, these authors observed
that the static barrier $\Delta E$ lies always inside the error
bars of $\Delta E^{TI}$ and $\Delta E^{MD}$. We will see in the
following that, similarly, our results on the diffusion of Cu
along straight steps of Cu(111) can be fitted nicely by assuming
$\Delta E^{MD}=\Delta E$.

\section{Diffusion of a Cu adatom along straight steps on Cu(111)}

\subsection {The geometry}

It is well known that there are two types of steps with
close-packed edges on Cu(111): the step A with a (100) ledge and
the step B presenting a ($\bar{1}11$) ledge. The study of the
surface diffusion along or across these steps is a prerequisite to
investigate the growth of large Cu adislands on Cu(111). These
adislands are expected to be bordered by steps of type A or B and,
out of equilibrium, may be or not in stacking fault\cite{Giesen03} since the
stacking fault energy is rather small in copper. In the first case
the atoms occupy HCP sites while, in the second, they are located
at FCC sites. Thus we have to study four geometries: step A or
step B for adislands at FCC or HCP sites.

\begin{figure}[!fht]
\begin{center}
\includegraphics*[scale=0.5,angle=0]{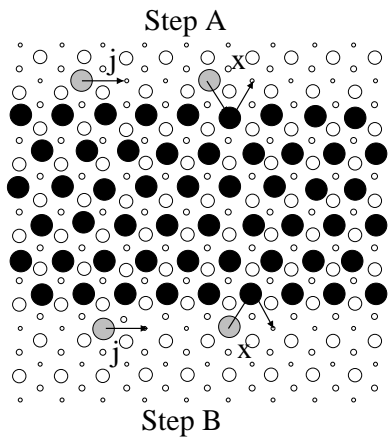}
\end{center}
\caption{Diffusion processes of a Cu adatom (gray circles) along
the steps A and B limiting a stripe in FCC geometry on Cu(111): $j$
(jump) and $x$ (exchange) mechanisms. The filled black circles are
the atoms of the stripe and the atoms of the substrate are denoted
as open circles with a size decreasing when going towards the inside
of the substrate.  }
\label{fig:jump_exch_AB}
\end{figure}

In practice a super-cell containing $p_1 \times p_2$ two
dimensional (2D) unit cells with $N_p$ layers is built to
represent the Cu(111) surface. On this super-cell a stripe of
$N_{stripe}$ close-packed atomic rows is added at FCC or HCP sites.
This stripe is thus limited by an A step on one side and a B step
on the other one (Fig.\ref{fig:jump_exch_AB}). The number
$N_{stripe}$ of rows is large
enough to avoid any interaction between the two steps. On each
side of this stripe (A and B edges) an adatom is deposited.
Finally, usual periodic boundary conditions are applied.

\subsection{Diffusion of adatoms in the TST-HA approach}

We have first carried out static calculations from which the
minimum energy path between two first neighbor adsorption sites
along A and B-steps is determined for the jump and exchange
diffusion mechanisms shown in Fig.\ref{fig:jump_exch_AB}. These
calculations were performed on a large super-cell: the slab
representing the surface contained ($11 \times 11$) 2D unit cells and was 10
layer thick. The stripe was made of 4 atomic rows. In view of the
large number of atoms in this super-cell, only the point
$\mathbf{k_{//}}=0$ in the surface Brillouin zone was used to calculate
the phonon frequencies.
Similarly to our previous work \cite{Marinica04}, the equilibrium
structure was deduced from the conjugate gradient method and
convergence was achieved when all the forces were smaller than
$10^{-3}$eV/\AA. The minimum energy path was determined using the
Ulitsky-Elber algorithm \cite{Ulitsky90} and the iteration process
was stopped when all the forces perpendicular to the path were
less than $3 10^{-2}$eV/\AA.

\begin{table}[!fht]
\begin{tabular}{|c||c|c|c|c|}
\hline
 Step type             &    \multicolumn{2}{c|}{A}   & \multicolumn{2}{c|}{B}    \\
       \hline
 Stripe geometry       &  FCC & HCP   &  FCC   &   HCP  \\
       \hline
$\Delta E_{j}$ (meV)   & 247  &  235   &  312  &  302       \\
 $\Gamma_0^{HA}=2\nu_0$(THz)  & 7.84 &  8.54  &  7.44 &  8.04       \\
$\Delta E_{e}$ (meV)   & 1532 &  1512  &  1715 &  1691      \\
\hline
\end{tabular}
\caption{Diffusion barrier $\Delta E_{j}$ and corresponding
prefactor $\Gamma_0^{HA}$ ($\nu_0$ is the Vineyard attempt
frequency) for Cu self-diffusion (jump mechanism) along A and
B-steps on the (111) surface, in regular FCC and faulted HCP geometries.
In the last line we have also reported the diffusion barrier
$\Delta E_{e}$ for the exchange mechanism shown in Fig.\ref{fig:jump_exch_AB}.}
\label{tab:diff_ad_barrier_step_A_andB}
\end{table}

\begin{figure}[!fht]
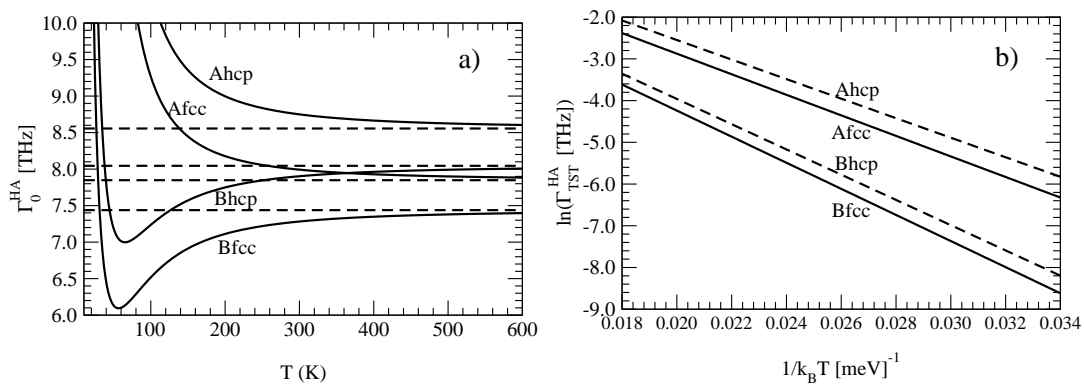

\begin{center}
\includegraphics*[scale=0.45,angle=0]{figure3a.eps}
\includegraphics*[scale=0.45,angle=0]{figure3b.eps}
\end{center}
\caption{a) Prefactor $\Gamma_0^{HA}(T)$ of the jump rate,
obtained from the TST-HA model, for Cu diffusion along A and
B-steps on Cu(111), in FCC and HCP geometries. b) Arrhenius plot
of $\Gamma_{TST}^{HA}(T)$ corresponding to the four geometries in
the temperature range $340K-645K$.} \label{fig:prefactor}
\end{figure}

\begin{figure}[!fht]
\begin{center}
\includegraphics*[scale=0.5,angle=0]{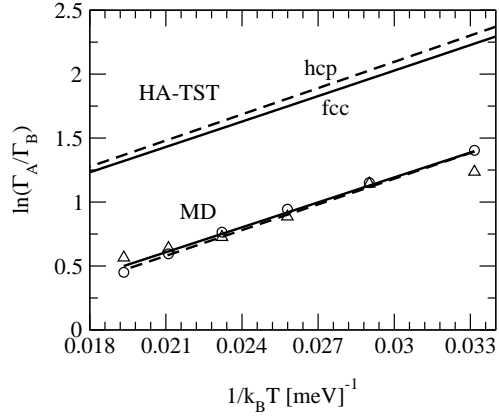}
\end{center}
\caption{Arrhenius plots of the ratio $\Gamma_A/\Gamma_B$ of the
jump diffusion rates for a Cu adatom along A and B steps on
Cu(111) in FCC and HCP geometries as deduced from the TST-HA model
and MD simulations (open circles: FCC geometry, open triangles: HCP geometry).} 
\label{fig:static_hopping_rate_AB_fcchcp}
\end{figure}

The results show that the adsorption energy of an adatom along a
B-step is favored by 7 meV with respect to an A-step.  This
difference in energy is probably too small to have any influence
on the island growth morphology and could not be at the origin of
an asymmetry between A and B-steps. In contrast
 the diffusion paths along step A and step B for
the jump mechanism are quite different. While along step A the
diffusion proceeds by FCC $\rightarrow$ HCP(saddle point)$\rightarrow$ FCC
jumps on the (111) terrace, along step B the path can be
approximated by a displacement along the channel of a (011)
microfacet which can be considered as the ledge. Indeed let us
recall that the (111)FCC surface with periodic monoatomic B steps
can be denoted either as (p+1)(111)$\times$($\bar{1}$11) or as
p(111)$\times$(011)\cite{Raouafi02b}. The resulting energy barriers
presented in Table \ref{tab:diff_ad_barrier_step_A_andB} show that
diffusion always proceeds by the jump mechanism (the energy
barrier for the exchange mechanism being extremely high) and is
expected to be faster along step A than along step B since its
energy barrier is smaller by about 60meV. Moreover, for both steps
the diffusion barrier along a HCP stripe is slightly smaller by
10meV than along a FCC stripe, i.e., the variation of the energy
barrier between A and B steps is almost exactly the same in FCC
and HCP geometries: 65meV and 67meV, respectively. Consequently
from static calculations, the asymmetry between the diffusion
along A and B-steps should be rather similar in FCC and HCP
geometries. Then we have calculated the phonon spectrum for
adatoms at the equilibrium and saddle point configurations, from
which the prefactor $\Gamma_0^{HA}(T)$ is deduced. In
Fig.\ref{fig:prefactor}a we show the evolution with temperature of
this prefactor: it is seen that the high temperature limit is
attained above $\simeq 300K$, i.e., close to the Debye temperature
of Cu. This limit is equal to twice the Vineyard attempt frequency
(Eq.\ref{nu0}) and is given in Table
\ref{tab:diff_ad_barrier_step_A_andB}. Accordingly, above 300K, 
the diffusion rate follows an almost
perfect Arrhenius law which is actually almost undistinguishable
from its asymptotic limit $2\nu_0 \exp(-\Delta E/k_BT)$
(Fig.\ref{fig:prefactor}b). One notes that the prefactors are not
very dependent on the geometry and thus, in this model, the
diffusion rate is governed almost entirely by the exponential
term. The diffusion along step A is indeed faster than along step
B and the motion is slightly easier in the HCP geometry. In order
to quantify the asymmetry between step A and step B we have also
calculated the ratio $\Gamma_A/\Gamma_B$ of the diffusion rates
along the two types of steps (see Fig.
\ref{fig:static_hopping_rate_AB_fcchcp}). This ratio is very
similar in HCP and FCC geometries, it increases as the temperature
decreases and typically adatom diffusion processes are
approximately 4 and 6 times faster along step A than along step B
at 600K and 450K, respectively.

\subsection{Diffusion of adatoms from MD simulations}

The simulations have been performed on a $(11 \times 11)$ slab
containing 16 layers with a stripe made of 4 atomic rows in FCC
sites on one side of the slab and in HCP sites on the other side.
One adatom is deposited along each of the four close-packed step
edges of the two opposite stripes. This procedure allows to obtain
with a single simulation box, the four trajectories from which the
diffusion coefficients are extracted. We have performed a series
of MD simulations for temperatures ranging from 350K to 600K. The
classical equations of motion were integrated with a time step of
3.5fs. The system was first equilibrated for about 12ps. In the
considered temperature range the far most frequent event is a jump
along the step, however from time to time an extra event occurs,
such as the formation of a dimer with an atom escaping from the
step (see Sec. \ref{other_events}). To avoid the occurrence of such ``unwanted
events'', we have preferred to perform  a number of independent
simulations rather than a single and longer one. At each
temperature the total simulation time was chosen in order to get a
meaningful statistics, i.e., it increases when the temperature
increases. The diffusion coefficient can be calculated either from
the Einstein relation (Eq.\ref{Dmd}), or from Eq.\ref{Djump} which
assumes uncorrelated discrete jumps. We have found that the two
methods give practically the same results showing that the jumps
are indeed uncorrelated. In the following the results will be
analyzed using the second approach which has the advantage of
allowing an estimation of statistical errors. Thus the analysis of
the trajectory consists in an enumeration of the various jumps
(accepted or rejected) occurring during the simulation time. The
statistics of the number of jumps is presented in Table
\ref{tab:jumps_statistics}. It is seen at first glance that
significant differences with TST-HA are expected since the
transmission coefficient $\kappa$ (i.e., the fraction
$N_1/(N_1+N_r)$ of accepted jumps) is far from being unity.

 \begin{table}[!fht]
\begin{tabular}{|c|c|cccc|cccc||cccc|cccc|}
\hline
\multicolumn{2}{|c|}{Step type}        &    \multicolumn{8}{c||}{A}   & \multicolumn{8}{c|}{B}    \\
       \hline
\multicolumn{2}{|c|}{Stripe geometry}  &  \multicolumn{4}{c|}{FCC}
& \multicolumn{4}{c||}{HCP}   &
  \multicolumn{4}{c|}{FCC}   &   \multicolumn{4}{c|}{HCP}  \\
 \hline
T(K) & t(ns)      & $N_1$&$N_2$&$N_3$&$N_r$&$N_1$ &$N_2$&$N_3$&$N_r$&$N_1$ &$N_2$&$N_3$&$N_r$&$N_1$ &$N_2$&$N_3$&$N_r$  \\
\hline
350 &  165.9      & 277  &  3  & 0   & 258 & 401  & 9   & 0   & 180 & 71   & 0   & 0   & 137 & 115  & 3   & 0   & 82 \\
400 &  98.0       & 391  & 17  & 1   & 382 & 608  & 19  & 1   & 343 & 127  & 3   & 1   & 259 & 193  & 7   & 0   & 187 \\
450 &  77.7       & 696  & 21  & 1   & 582 & 953  & 30  & 2   & 553 & 275  & 8   & 0   & 478 & 377  & 16  & 1   & 384  \\
500 &  58.8       & 960  & 33  & 3   & 875 & 1294 & 38  & 3   & 727 & 447  & 19  & 0   & 779 & 544  & 31  & 5   & 561  \\
550 &  39.2       & 977  & 55  & 4   & 993 & 1323 & 60  & 3   & 815 & 540  & 26  & 4   & 842 & 603  & 50  & 4   & 722  \\
600 &  30.8       & 1159 & 57  & 7   & 1020& 1448 & 88  & 9   & 926 & 709  & 38  & 7   & 1249& 769  & 57  & 8   & 910  \\
\hline
\end{tabular}
\caption{Statistics of single ($N_1$), double ($N_2$), triple
($N_3$) and rejected single ($N_r$) jumps for the diffusion of a Cu adatom along
step A and step B on Cu(111), in FCC and HCP geometries, from MD
simulations. The total simulation time t is given for each
temperature T.} \label{tab:jumps_statistics}
\end{table}

\begin{figure}[!fht]
\begin{center}
\includegraphics*[scale=0.5,angle=0]{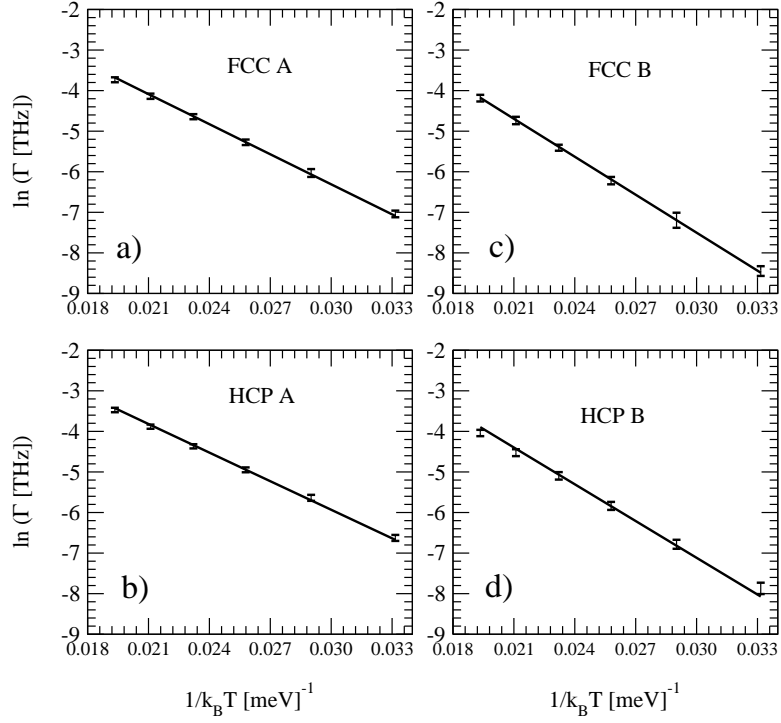}
\end{center}
\caption{Arrhenius plots of the MD jump rates for Cu
diffusion along straight steps on Cu(111) in FCC and HCP geometries.
The error bars give the statistical error.
The straight line is a least mean square fit in which the energy
barrier has been fixed to its static value. a)Step A in FCC
geometry; b)Step A in HCP geometry; c)Step B in FCC
geometry; d)Step B in HCP geometry.}
\label{fig:dynamic_hopping_rate_AB_fcchcp}
\end{figure}

In Fig.\ref{fig:dynamic_hopping_rate_AB_fcchcp} we show the
Arrhenius plots of $\Gamma^{MD}(T)$ for the four geometries
with error bars corresponding to the standard deviation. We have
first carried out a least mean square fit with the two parameters
$\Gamma_{0}^{MD}$ and $\Delta E^{MD}$. Similarly to the results of
Boisvert {\it et al.} \cite{Boisvert98} on the surface diffusion
of Cu on Cu(100), we find that $\Delta E^{MD}$ is quite close to
the static barrier $\Delta E$ (see Table \ref{tab:MDfit_results}) and the fitted straight
line lies inside the error bars demonstrating the quality of the
statistics. However, as already discussed in Sec.IIB2, it is
advisable to compare with a fit in which the barrier has been
fixed to its static value. The corresponding fits shown in Fig.
\ref{fig:dynamic_hopping_rate_AB_fcchcp} are nearly as good as the
previous ones. Moreover the associated prefactors become quite
comparable in FCC and HCP geometries in contrast with the previous fits
for which the prefactor for the B step in the HCP geometry was much smaller
than in the FCC geometry. This does not seem physically reasonable
but is most probably due to the uncertainty on $\Delta E^{MD}$
since a small error in $\Delta E^{MD}$ must be compensated by a
large variation of the prefactor in order to give the same value
of $\Gamma^{MD}$. Thus we have adopted the values of the
prefactors $\Gamma_{0}^{MD}$ derived from the fit in which $\Delta
E^{MD}=\Delta E$. In this case it is seen from Tables
\ref{tab:diff_ad_barrier_step_A_andB} and \ref{tab:MDfit_results}
that $\Gamma_{A}/\Gamma_{B}$ as deduced from MD is about twice
smaller than the corresponding value in TST-HA
(see Fig.\ref{fig:static_hopping_rate_AB_fcchcp}). Indeed
in the TST-HA model the prefactors are very similar for steps A
and B while, from MD simulations, the prefactor corresponding to
steps B, which have the highest diffusion barrier, is about twice that
of step A. We will now give a physical interpretation for this
difference.

\begin{table}[!fht]
\begin{tabular}{|c|c||c|c|c|c|}
\hline
 \multicolumn{2}{|c||}{Step type}             &    \multicolumn{2}{c|}{A}   & \multicolumn{2}{c|}{B}    \\
       \hline
\multicolumn{2}{|c||}{Stripe geometry}       &  FCC & HCP   &  FCC   &   HCP  \\
       \hline
\multicolumn{2}{|c||}{$<\kappa>$}            & 0.52 & 0.64  & 0.36   &  0.50  \\
 \hline
(a) &$\Delta E^{MD}$ (meV) fitted   & 239  &  226   &  308  &  279       \\
  & $\Gamma_0^{MD}$(THz) fitted     & 5.0  &  4.9   &  11.6 &  7.8       \\
\hline
(b) &$\Delta E$ (meV) fixed        & 247  &  235    &  312  &  302       \\
  & $\Gamma_0^{MD}$(THz) fitted    & 6.0  &  6.1    &  12.8 &  14.0       \\
\hline
\end{tabular}
\caption{Transmission coefficients $<\kappa>$ averaged over
temperature, energy barriers and prefactors derived from MD
simulations: a)both parameters $\Delta E^{MD}$ and $\Gamma_0^{MD}$
have been obtained from a least mean square fit; b)$\Gamma_0^{MD}$
has been obtained from a least mean square fit with $\Delta E^{MD}$
fixed at its static value $\Delta E$.} \label{tab:MDfit_results}
\end{table}

It is frequently found that when the activation energy increases
within a family of processes, the prefactor also increases. Thus,
in the expression of the process rate, the increase of the
prefactor somewhat ``compensates'' for the decrease in the Arrhenius
exponential term governing the dependence on temperature. This effect is known as the
Meyer-Neldel compensation law \cite{Meyer37}. It has been
investigated by Boisvert {\it et al.} \cite{Boisvert95} in the
particular case of surface self-diffusion by means of MD
simulations. By studying jump and exchange processes on the (100)
and (111) surfaces of elements belonging to the end of the
transition series or to noble metals, they have proposed the
following law for the prefactors:

\begin{equation}
\Gamma_0=\Gamma_{00}\exp(\Delta E/\Delta_0)^{\alpha}
\label{Eq:MN}
\end{equation}

 From theoretical models \cite{Emin74,Yelon92} the exponent $\alpha$ is
 expected to lie in the range 0.5-1 depending on the nature of the
 excitations that give rise to the activated process. For
 acoustical phonons, a simple phenomenological model predicts
 $\alpha=3/4$ and a characteristic energy $\Delta_0$ of the order
 of 2-3 times larger than a typical phonon energy (for instance
 $h\nu=k_BT_D$ where $T_D$ is the Debye temperature). Actually Boisvert
 {\it et al.} \cite{Boisvert95} have found that their results can
 be reasonably fitted with $\alpha=0.7, \Delta_0=74$meV and
 $\Gamma_{00}=0.74$THz.

\begin{figure}[!fht]
\begin{center}
\includegraphics*[scale=0.5,angle=0]{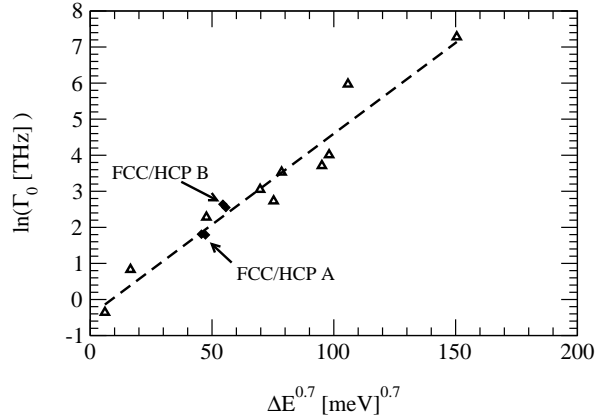}
\end{center}
\caption{Plot of $\ln\Gamma_0$ as a function of $\Delta E^{0.7}$.
The points represented by empty triangles are taken from Boisvert
{\it et al.} \cite{Boisvert95} and refer to various surface
self diffusion processes on low index surfaces of Au, Ag, Ni and Pd.
The points indicated by arrows (filled black diamonds) are derived
from our molecular dynamic simulations along A and B straight
steps on Cu(111) in FCC and HCP geometries. The dashed line is the fit
corresponding to Eq.\ref{Eq:MN}.} \label{fig:meyerneldel}
\end{figure}

 As already stated, our MD data for steps A and B obey the same
 trend since $\Gamma_0^{MD}$ increases with $\Delta E$. Thus it is
 interesting to add the corresponding data (see Fig.\ref{fig:meyerneldel})
 on the plot giving the logarithm of the prefactor versus $(\Delta E)^{0.7}$ which should
 be a straight line (Fig.3 of Ref.\onlinecite{Boisvert95}).
 Our results are quite close to the straight
 line obtained by Boisvert {\it et al.}.

 Let us now comment on the differences between the TST-HA and MD
 approaches. As already mentioned, MD simulations take into
 account the transmission coefficient $\kappa$ and anharmonic
 effects. From our simulations $\kappa$ does not vary
 significantly with temperature. From Table \ref{tab:MDfit_results}, it is seen that on
 average $\kappa_A/\kappa_B\simeq 1.5$ for both geometries. As a
 consequence the increase of the prefactor for step B 
 with respect to step A is not due to the
 transmission coefficient but rather to anharmonic effects which,
 as discussed by Boisvert{\it et al.}~\cite{Boisvert95,Boisvert98},
 lead to multiphononic excitations. The prefactor is then
 proportional to the number of ways of assembling these
 excitations. This gives rise to an entropy factor increasing with
 the barrier height (see Appendix). This interpretation is
 consistent with the absence of a Meyer-Neldel effect in the TST-HA
 model which has already been noted by other authors
 \cite{Ratsch98}.

 As a conclusion, it is clear that the prefactor should increase with
 the (static) barrier height. Moreover the law proposed by Boisvert~{\it et
 al.} (Eq.\ref{Eq:MN}) can be used to get a reasonable estimate of
 this variation. Thus, in the remaining part of this work, we will
 limit ourselves to the determination of the static diffusion barriers.

\section{Other diffusion events in the presence of straight steps}

\subsection{Diffusion of adatoms across straight steps: Schwoebel barrier}

\begin{figure}[!fht]
\begin{center}
\includegraphics*[scale=0.5,angle=0]{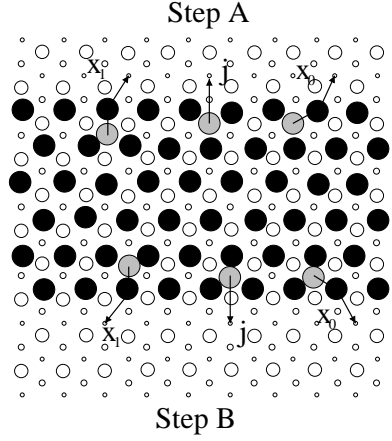}
\end{center}
\caption{Diffusion processes of a descending adatom across the A
and B- steps limiting a stripe in FCC geometry on Cu(111), by jump
(j) or exchange ($x_0$ and $x_1$) mechanisms. The atoms are
represented as in Fig.\ref{fig:jump_exch_AB}. }
\label{fig:schwobel}
\end{figure}

When an adatom is deposited on an adisland it wanders on its
surface until either it meets a group of adatoms already present
on this adisland or it comes down the step edge by diffusing
across the step. However the additional barrier (called the
Schwoebel barrier) felt by the adatom for the latter process is
often rather high relative to the barrier (40meV) encountered on
the flat surface and the probability that an adatom crosses the
step is small. As a consequence it is often assumed that the flux
of adatoms towards the step comes from the lower terrace. The
``traditional picture'' of an adatom ``jumping'' down  the step (process j
of Fig. \ref{fig:schwobel}) and therefore loosing one
nearest neighbor in the diffusion process usually leads to high
energy barriers but another type of diffusion process exists,
i.e., the exchange mechanism in which the adatom takes the place
of a step atom which becomes an adatom attached to the step
(processes $x_0$ and $x_1$ of Fig. \ref{fig:schwobel}). In Table
\ref{tab:schwobel_barrier} we present the energy barriers
corresponding to the three types of diffusion mechanisms across
the steps A and B on an adisland in FCC and HCP configurations.
Interestingly the energy barriers for a jump mechanism are almost
the same (around 510meV) in all cases, but the exchange mechanism
has a lower barrier by $\simeq$ 200meV for the process $x_0$
($\simeq$ 130meV for $x_1$) on step A and $\simeq$ 400meV for both
exchange processes on step B. A strong anisotropy is therefore
found between step A and step B. The Schwoebel barrier is thus
only $\simeq$ 50meV for step B while it is around 250meV and
300meV for $x_0$ and $x_1$, respectively, on step A. The existence
of a low Schwoebel barrier across step B is in agreement with the
previous EAM calculations by Trushin {\sl et al}\cite{Trushin97}
and could be at the origin the fast decay of double layer Cu
adislands on Cu(111) observed in
experiments\cite{Giesen99,Gong04}.

\begin{table}[!fht]
\begin{tabular}{|c||ccc|ccc||ccc|ccc|}
\hline
 Step     &  \multicolumn{6}{c||}{A}           &  \multicolumn{6}{c|}{B}  \\
       \hline
Geometry  & \multicolumn{3}{c|}{FCC}    &  \multicolumn{3}{c||}{HCP}   & \multicolumn{3}{c|}{FCC} & \multicolumn{3}{c|}{HCP} \\
\hline
Direction &  $j$ & $x_0$ & $x_1$ & $j$ & $x_0$ & $x_1$ & $j$ & $x_0$ & $x_1$ & $j$ & $x_0$ & $x_1$    \\
\hline
 $\btr$    &   507    &  299     &  374    &  508   &  289     & 362   &  510   &  90   & 106 & 510  & 86   & 103    \\
 $\btl$    &  1222    & 1014     &  1032   &  1218  &  999     & 1016  &  1228  &  807  & 776 & 1222 & 799  & 767     \\
       \hline
\end{tabular}
\caption{Diffusion barriers of an adatom descending or ascending a
step (A or B) by various mechanisms: jump ($j$), exchange ($x_0$
or $x_1$) (see Fig. \ref{fig:schwobel})for FCC and
HCP geometries. The directions $\btr$ and $\btl$ correspond to descending and
ascending adatoms, respectively.} \label{tab:schwobel_barrier}
\end{table}

\subsection{Formation and diffusion of dimers}

\begin{figure}[!fht]
\begin{center}
\includegraphics*[scale=0.5,angle=0]{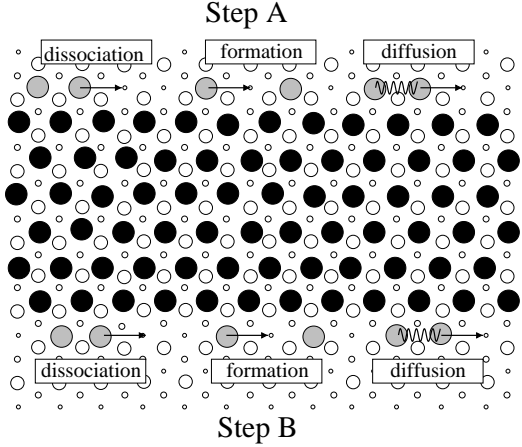}
\end{center}
\caption{Atomic motions corresponding to the dissociation, the
formation and the concerted diffusion of a Cu dimer along the steps A and B
limiting a stripe in FCC geometry on Cu(111). The atoms are
represented as in Fig.\ref{fig:jump_exch_AB}.}
\label{fig:dimer_diffusion}
\end{figure}

A step adatom diffusing along the step will possibly meet another
single adatom to form a step dimer, or stick to a kink. The latter
case will be discussed later (see Sec.V). Let us first mention
that the adsorption energy of a dimer along a B-step is favored by
5meV with respect to an A-step. In Table \ref{tab:dimer_formation}
we present the energy barriers for the formation and dissociation
of a dimer (see Fig.\ref{fig:dimer_diffusion}).  As expected the diffusion of an adatom is favored by
the proximity of another adatom, the energy barrier for the dimer
formation being around 15 \% lower than the energy barrier to
diffuse freely along a step. Once the dimer is formed the
probability for the dimer dissociation is rather low since in any
case the associated barrier is larger than 500meV (see Table
\ref{tab:dimer_formation}), however a concerted motion of the two
atoms is conceivable (see Fig.\ref{fig:dimer_diffusion}).

 \begin{table}[!fht]
\begin{tabular}{|c||c|c||c|c|}
\hline
 Step                  &         \multicolumn{2}{c||}{A}          &   \multicolumn{2}{c|}{B}  \\
            \hline
 Geometry               & FCC     &  HCP    &  FCC    &    HCP        \\
\hline
Dimer dissociation      &  520     &  509     &  568    &   560      \\
Dimer formation         &  207     &  197     &  262    &   255 \\
\hline
\end{tabular}
\caption{Energy barriers for the formation and  dissociation of a
step dimer along A and B-steps in FCC and HCP geometries.}
\label{tab:dimer_formation}
\end{table}

We have therefore calculated the energy barrier for the diffusion
of such step dimers (Table \ref{tab:dimer_diffusion}).
Interestingly the barrier height for the dimer motion shows a more
pronounced anisotropy between step A and step B than for the
single adatom motion. Indeed the ratio of the diffusion barriers
along steps A and B is $\simeq$ 1.27 and $\simeq$ 1.40 for the
single adatom and concerted dimer motions, respectively. In
addition the comparison of Tables \ref{tab:dimer_formation} and
\ref{tab:dimer_diffusion} shows that the concerted motion of the
dimer is easier than the dimer dissociation along step A, while
along step B the dissociation is favored. As a consequence, the
dimer diffusion will proceed by concerted motion along step A,
whereas along step B a two step process is preferred, i.e., a
dissociation followed by the re-bonding of the dimer.

\begin{table}[!fht]
\begin{tabular}{|c||c|c||c|c|}
\hline
 Step        &    \multicolumn{2}{c||}{A}  & \multicolumn{2}{c|}{B}   \\
       \hline
 Geometry    &  FCC & HCP   &  FCC   &   HCP  \\
       \hline
$E_{dim}$ (meV)   & 463   & 440   &  640   &  625       \\
\hline
\end{tabular}
\caption{Diffusion barriers $E_{dim}$ for the dimer diffusion
along steps A and B in FCC and HCP geometries.}
\label{tab:dimer_diffusion}
\end{table}

\section{Diffusion of adatoms along steps with defects}

When an adatom sticks onto a preformed close-packed adisland
bordered by A and B steps with defects (kinks or other adatoms) it
will in general remain attached to it and, if edge diffusion and
corner crossing processes are activated, the adatom will diffuse
back and forth between the two types of steps. In the following
section we will start by presenting the potential energy profiles
for typical diffusion sequences along which the most important
processes occur. Then the other diffusion processes will be
systematically investigated and the result for their activation
barrier will be given and commented on.

\subsection{Typical diffusion sequences}

\begin{figure}[!fht]
\begin{center}
\includegraphics*[scale=0.5,angle=0]{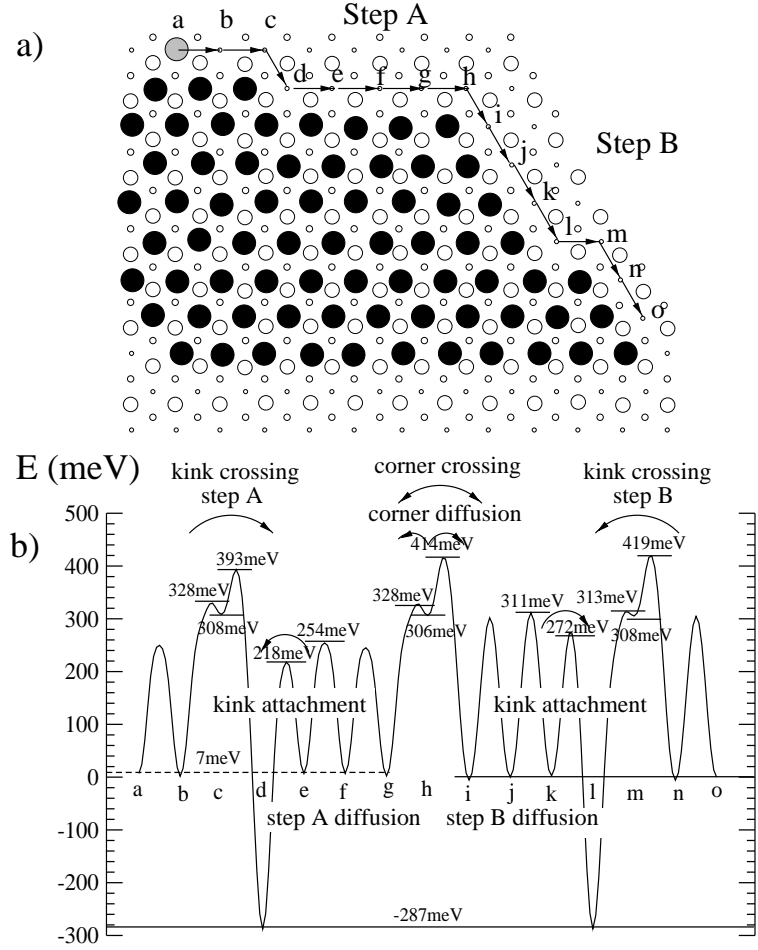}
\end{center}
\caption{a) A typical diffusion path of an adatom along the
borders of an adisland in FCC geometry with A and B edges on Cu(111), b) The
associated energy profile. The atoms are represented as in
Fig.\ref{fig:jump_exch_AB}.} \label{fig:diffusionpath}
\end{figure}

The diffusion path around an adisland in FCC geometry with A and (or) B 
borders and the corresponding energy profile is shown in
Fig.\ref{fig:diffusionpath}. An adatom starting from step A passes
around a kink (kink crossing), detaches from this kink, diffuses
along step A and then passes around a corner between step A and
step B (corner crossing) on which a similar path is followed. Such
a sequence is very instructive since the most important processes
are encountered (the formation and dissociation of dimers along
the steps have been studied in Sec.IV.B): \textit{i)} diffusion
along step A ($a\leftrightarrows b, e\leftrightarrows f$)and along
step B ($j\leftrightarrows k, n\leftrightarrows o$), \textit{ii)}
kink crossing on step A ($b\rightarrow c \rightarrow d)$ and on
step B ($n\rightarrow m \rightarrow l)$, \textit{iii)} kink
attachment ($e\rightarrow d$ on step A, $k\rightarrow l$ on step
B) or detachment ($d\rightarrow e$ on step A, $l\rightarrow k$ on
step B), \textit{iv)} corner crossing ($g\leftrightarrows h
\leftrightarrows i$).

The energy barriers for the diffusion along step A and step B have
been already studied (Sec.III.B). The numerical values given in
Fig.\ref{fig:diffusionpath}b are very slightly different (by a few
meV) due to the finite size of the adisland edges.

The kink crossing is a two step process since a secondary energy
minimum is found at site $c$ on step A and $m$ on step B. The
energy barrier to pass around the kink is significantly larger
than for the diffusion along straight steps. This is actually the
one-dimensional analogue of the Schwoebel effect encountered when
diffusing across steps. This means that the adatom is repelled by
the descending kink. Furthermore even if the adatom reaches site
$c$ on step A, it will most probably come back to site $b$
(barrier: 20meV) than continue its way to site $d$ (barrier:
85meV). On the contrary on step B an adatom reaching the site $m$
will most probably continue its way to site $l$ (barrier: 5meV)
rather than turning back to site $n$ (barrier: 111meV).

The energy barriers for kink detachment are rather high (505meV
for step A, 559meV for step B) since roughly two nearest neighbor
bonds have been broken at the saddle point. Moreover the energy
barriers for the kink attachment are significantly lowered (by
30-40meV) compared with the diffusion barriers of the straight
steps. Thus an adatom reaching site $e$ or $k$ has a larger
probability to attach to the ascending kink than to be reflected.
It is worthwhile to note that this effect cannot be obtained
when using approximate expressions for the barriers.
Finally it is interesting to compare these energy barriers to
those involved in the dissociation and formation of dimers along
steps A and B. By comparing with the results of Table V, we see
that the kink detachment is easier than the dimer dissociation,
while the kink attachment is more difficult than dimer formation.
This is clearly due to the change of local atomic environment.

The energy profile for the corner crossing has some similarity
with the kink crossing since the energy barrier is significantly
larger than for the diffusion along straight steps, i.e., the
adatom is repelled by the corner leading to a Schwoebel-like
effect. Furthermore a secondary energy minimum is met at site $h$
during the process. However the two minima on both sides of site
$h$ (sites $g$ and $i$) are roughly at the same energy (the energy
difference between sites $g$ and $i$ being $\simeq 7meV$) since
both sites have the same first nearest neighbor coordination.
Moreover it is interesting to note that an adatom located at site
$h$ has a larger probability to escape towards step A (barrier:
22meV) than towards step B (barrier: 108meV).

\begin{figure}[!fht]
\begin{center}
\includegraphics*[scale=0.5,angle=0]{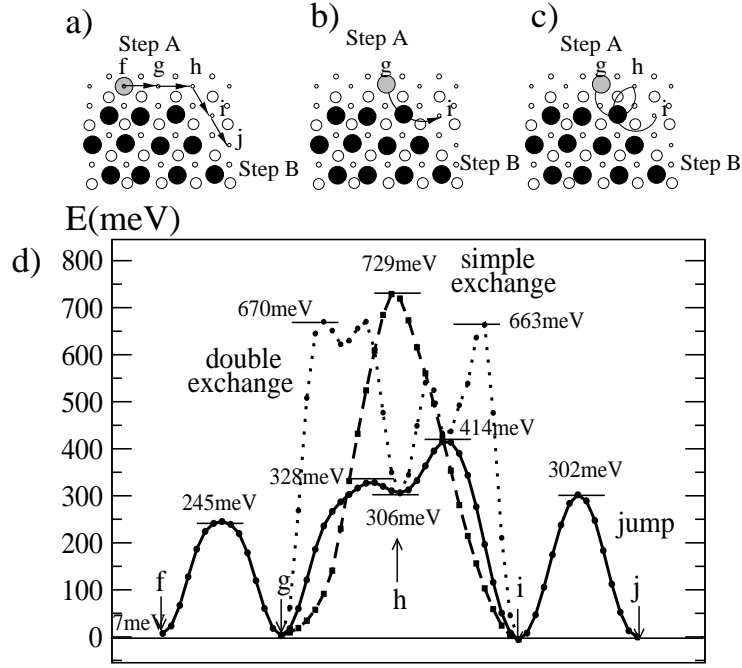}
\end{center}
\caption{Possible diffusion mechanisms for passing around the
corner between step A and step B on an adisland in FCC
geometry on Cu(111): a)jump process, b)single
exchange and c)double exchange mechanisms, d)the corresponding
potential energy profiles: jump (full line), single exchange
(dashed line) and double exchange (dotted line).The atoms are
represented as in Fig.\ref{fig:jump_exch_AB}. }
\label{fig:excorn}
\end{figure}

In addition, contrary to the diffusion across the straight steps
(see Table IV) the exchange mechanism is never favored compared
with the jump process. Indeed, we have already seen (Table I) that
the diffusion along steps does not proceed by exchange. Similarly
we have verified that the exchange mechanism is also unfavorable
for kink and corner crossings. This is illustrated in Fig.
\ref{fig:excorn} for the corner crossing: the usual double jump
mechanism ($g\rightarrow h \rightarrow i$) is much preferred to a
single (Fig.\ref{fig:excorn}b) or double exchange
(Fig.\ref{fig:excorn}c) mechanism. Surprisingly however the rather
unexpected double exchange mechanism is less costly than the
single exchange.

\begin{figure}[!fht]
\begin{center}
\includegraphics*[scale=0.6,angle=0]{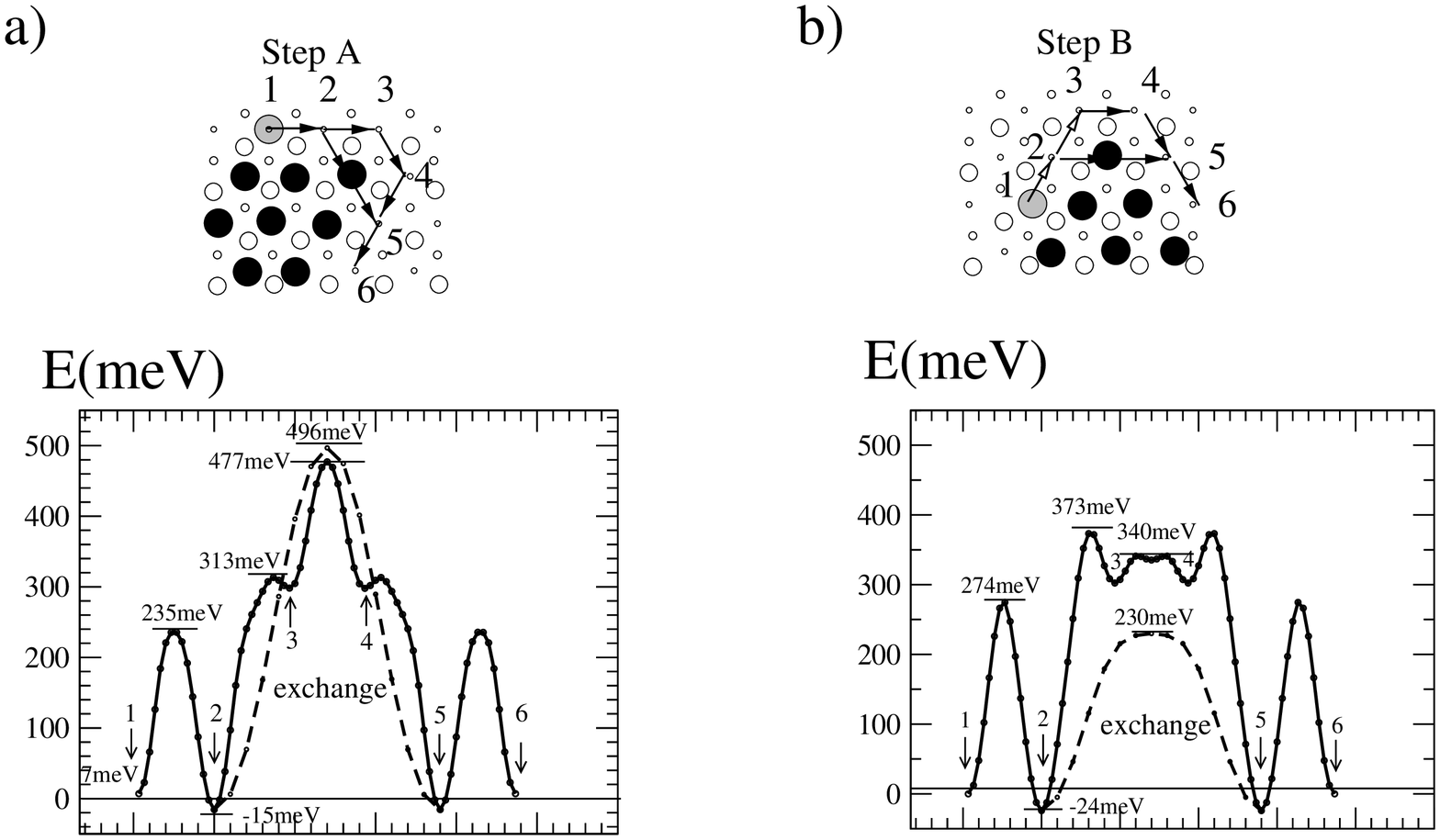}
\end{center}
\caption{Upper part: possible diffusion mechanisms (jump and exchange) for
passing around the corner of a triangular adisland bordered a) by
A-steps or b) B-steps in FCC geometry on Cu(111). The atoms are
represented as in Fig.\ref{fig:jump_exch_AB}. Lower part: the corresponding
potential energy profiles: jump (full line) and exchange (dashed
line) mechanisms. } \label{fig:excorn2}
\end{figure}

Bearing in mind  that crystal growth is a kinetic phenomenon
leading to out of equilibrium shapes like triangular adislands, it
is also important to study typical diffusion processes occurring
in the vicinity of such adisland shapes. For this reason we have
also considered the corner crossing of a triangular adisland
bordered with A or B steps. The results of our calculation,
illustrated in Fig. \ref{fig:excorn2}, is quite instructive since
it is found that the corner crossing of a triangular adisland
bordered with B steps is much easier than for a triangle with A
edges. This asymmetry already exists if the crossing mechanism
occurs by successive elementary jumps, but it is much more
pronounced for the exchange mechanism which is highly favorable at
the corner of a triangular adisland bordered with B edges. Indeed
the corresponding activation energy is only 230meV, i.e., even
much less than the simple diffusion along the straight step B.
This could have important consequences on the growth scenario.

\subsection{Other diffusion events \label{other_events}}

\begin{figure}[!fht]
\begin{center}
\includegraphics*[scale=0.5,angle=0]{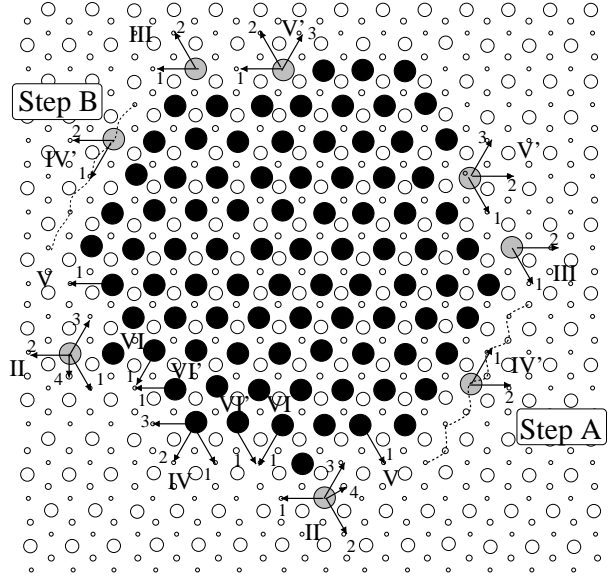}
\end{center}
\caption{Diffusion processes of an adatom around an
adisland bordered with A and B steps at FCC sites on Cu(111).
The atoms are
represented as in Fig.\ref{fig:jump_exch_AB}.} \label{fig:all_processes}
\end{figure}

Finally, in view of carrying out a KMC simulation
of Cu/Cu(111) growth it is important to have a complete picture of
the various diffusion processes. An enumeration of (almost) all
diffusion mechanisms (labelled by roman figures) of an adatom in
the vicinity of a preformed adisland is presented in Fig.
\ref{fig:all_processes}.  For the sake of completeness we have
also considered the possibility for a step edge atom ($V$,
$VI$, $VI'$) to escape along the step  or a step adatom ($IV'$) to
escape on the terrace. We use the following notation hereafter:
the initial configuration of the diffusion events represented in
Fig.\ref{fig:all_processes} is denoted by an index 0 and the final
configuration by an index $i=1,2,3,4$. For example the simple
diffusion of an adatom along a straight step A is denoted as:
$\text{IV'}_{0\rightarrow 1}^{A}$. The arrow is simply reversed
for the opposite displacement. Note that, even though FCC-HCP
jumps are the smallest possible diffusion movements, they were not
taken into consideration here (except for $\text{II}_{0\rightarrow 4}^{A}$
discussed in the following) since in the vicinity of an adisland
all the sites that can be reached by such jumps are
unstable, and an adatom situated on such a site
would immediately be attracted by the step.

Most of the diffusion processes along steps have already been
discussed in detail above and we shall not make any further
comment on them. Let us however try to extract some general trends
from Table \ref{tab:all_processes}.

One can observe that  the diffusion energy of an atom in the
vicinity of an adisland in HCP geometry is systematically smaller
by a few meV than the corresponding energy in the vicinity of an
adisland in FCC geometry similarly to the diffusion along straight
steps.

Obviously, the energy necessary  for an atom to leave an adisland
towards the terrace and loose all its first nearest neighbors from
the adisland, is much higher than the diffusion energy along
steps. Indeed the energy cost of such processes is around 950meV
for a corner ($\text{IV}_{0\rightarrow 2}$) or a kink atom
($\text{V'}_{0 \rightarrow 2}$), around 650meV for a step adatom
to escape on the terrace ($\text{IV'}_{0\rightarrow 2}$) and
decreases to 400meV for the motion $\text{II}_{0 \rightarrow 2}^B
$. These energy barriers obviously decrease with the number of
bonds lost in the diffusion process which are equal to 3, 2 and 1,
respectively. However this correlation is very approximate since,
for a given number of broken bonds, the values of the activation
barriers vary over a rather large interval: for instance, when a
single bond is broken the activation barrier lies between 323 and
560meV. Indeed, similarly to the case of straight steps, the
potential energy surface of an adatom is strongly affected by the
vicinity of the adisland. As a consequence the diffusion path
cannot be guessed from symmetry arguments and, at the saddle
point, many different interatomic distances with the neighbors are
involved so that it is difficult to derive an effective pair
interaction model as done in our previous work\cite{Marinica04}.
Let us also stress that the terrace site that can be reached from
an adisland site may be unstable or quasi unstable, see for
example the diffusion processes $\text{IV}_{0\leftarrow 2}$ and
$\text{V'}_{0\leftarrow 2}$ for both steps.

Finally let us mention that the $\text{II}_{0\rightarrow 4}^{A}$
process is also affected by the presence of the adisland. Indeed
the energy profile would be very different if the diffusing atom
and its nearest neighbor were isolated on the terrace. The dimer
motion was studied in detail in our previous work
\cite{Marinica04} and the energy barrier corresponding to  a
motion of type $\text{II}_{0\rightarrow 4}$ from an FCC-FCC to an
FCC-HCP configuration (also called ff-fh  in
Refs.\onlinecite{Marinica04,Repp03}) was found equal to 16meV (the
initial and final sites being almost energetically degenerate),
whereas in the presence of an adisland this motion becomes highly
asymmetrical, and the true dimer (FCC-FCC) is stabilized by the
vicinity of the adisland. This phenomenon is similar with the
experimental finding of Repp {\it et al.}\cite{Repp03} who observed the
stabilization  of a dimer by the proximity of a monomer. In the
case of $\text{II}_{0\rightarrow 4}^{B}$ the influence of the step atoms is
so strong that the final site becomes unstable.

\begin{table}[!fht]
\begin{tabular}{|cc|c||c|c||c|c|}
\hline
           &       &                 &  \multicolumn{2}{c}{A}          &   \multicolumn{2}{c|}{B}  \\
       \hline
\multicolumn{2}{|c|}{Diffusion event}    & $\Delta Z$  & FCC     &  HCP    &  FCC    &    HCP        \\
\hline
II         & $0 \rightarrow 1$  &  0    & 38      &  33     &   189   &   180      \\
           & $0 \leftarrow 1$   &       & 38      &  33     &   189   &   180      \\
\hline
           & $0 \rightarrow 2$  &  1    & 400     &  396    &   411   &   405        \\
           & $0 \leftarrow 2$   &       & 30      &  28     &   32    &   29        \\
\hline
           & $0 \rightarrow 3$  &       & 60      &  54     &  3      &   3        \\
           & $0 \leftarrow 3$   &  2    & 668     &  660    &  600    &   602      \\
\hline
           & $0 \rightarrow 4$  &       & 39      &  33     &  -      &    -       \\
           & $0 \leftarrow 4$   &       &  6      &  10     &  -      &    -       \\
\hline \hline
III        & $0 \rightarrow 1$  &  1    & 327     &  323    &   424   &   415      \\
           & $0 \leftarrow 1$   &       & 21      &  17     &   113   &   104       \\
\hline
           & $0 \rightarrow 2$  &  2    & 667     &  664    &   689   &   686       \\
           & $0 \leftarrow 2$   &       & 3       &  1      &    17   &   16         \\
\hline \hline
IV         & $0 \rightarrow 1$  &  2    &  609    &  606    &  609    & 606        \\
           & $0 \leftarrow 1$   &       &  10     &  9      &  10     & 9        \\
\hline
           & $0 \rightarrow 2$  &  3    &  952    &  948    &  952    &  948          \\
           & $0 \leftarrow 2$   &       &   0     &   0     &   0     &  0      \\
\hline
           & $0 \rightarrow 3$  &  2    &  695   &  685     &  695    &  685        \\
           & $0 \leftarrow 3$   &       &  94    &  87      &   94    &  87          \\
\hline \hline
IV'        & $0 \rightarrow 1$  &  0    &  247    &  235   &  312    &  302         \\
           & $0 \leftarrow 1$   &       &  247    &  235   &  312    &  302          \\
\hline
           & $0 \rightarrow 2$  &  2    &  660    &  661    &  683    &  677          \\
           & $0 \leftarrow 2$   &       &  0.6    &  0.7    &  19     &  15          \\
\hline \hline
V          & $0 \rightarrow 1$  &  3    &  913    &  903    &  866    &  862           \\
           & $0 \leftarrow 1$   &       &  45     &  40     &   0     &  0            \\
\hline \hline
V'         & $0 \rightarrow 1$  &  1    &  506    &  494    &  560  &  551           \\
           & $0 \leftarrow 1$   &       &  211    &  200    &  270  &  262            \\
\hline
           & $0 \rightarrow 2$  &  3    &  950    &  947    &  947    &  945             \\
           & $0 \leftarrow 2$   &       &   0     &  0      &   0     &  0           \\
\hline
           & $0 \rightarrow 3$  &  2    &  681    &  674    &  600    &  599            \\
           & $0 \leftarrow 3$   &       &  86     &  80     &   8     &  7         \\
\hline \hline
VI         & $0 \rightarrow 1$  &  3    & 842     &  837    & 824     &   816          \\
           & $0 \leftarrow 1$   &       &  45     &   43    &  19     &    16          \\
\hline \hline
VI'        & $0 \rightarrow 1$  &  2    & 726     &  713    &  647    &    633         \\
           & $0 \leftarrow 1$   &       & 196     &  183    &  114    &    103          \\
\hline
\end{tabular}
\caption{Calculated diffusion barriers for a Cu adatom jumping
from an initial site 0 to a final site (1, 2, 3, or 4) and
vice-versa. $\Delta Z$ is the number of nearest neighbor bonds
broken in the diffusion process. The various diffusion events are
shown in Fig.\ref{fig:all_processes}.} \label{tab:all_processes}
\end{table}

\section{Conclusion}

In conclusion, we have carried out a systematic study of the
diffusion processes that can occur for a Cu adatom in the presence
of a close-packed adisland on Cu(111). In view of the small value
of the stacking fault energy in Cu, two geometries have been
considered in which the atoms of the adisland occupy FCC or HCP
sites. The diffusion rates along step A and step B without defects
have been first investigated using TST-HA and MD simulations in
the classical limit. They obey Arrhenius laws with activation
barriers significantly larger along step B than along step A.
Whereas the (static) potential energy barriers account for the
slope of the MD Arrhenius plots within statistical errors, the
attempt frequencies are markedly different due to the anharmonic
and recrossing effects included in MD simulations. Indeed, in
contrast with the Vineyard attempt frequencies which have similar
values for all geometries, the prefactors derived from MD
simulations obey the Meyer-Neldel compensation rule, i.e., the
increase of the activation barrier for B steps compared with A steps is somewhat
compensated by an increase of the prefactor. Furthermore, the law
proposed by Boisvert {\it et al.} \cite{Boisvert95} to relate the
activation barrier to the prefactor accounts quite well for our
results. Consequently we have then limited ourselves to the
determination of static potential barriers for a large number of
diffusion events that can occur in the presence of an adisland. A
number of additional differences have been put forward between the
diffusion along A and B borders which may have an influence on the
adisland growth shapes. In agreement with previous EAM
calculations \cite{Trushin97} we find that the exchange mechanism
for descending a step is largely favored, especially on step B for
which the Schwoebel extra barrier is only $\simeq$50-60meV.
Moreover it is shown that the corner crossing diffusion process by
jump or exchange for triangular adislands with A borders have a
similar and high ($\simeq$480meV) activation barriers while for
triangular adislands with B borders the exchange mechanism is
largely favored and has a rather low barrier (230meV). From this
set of results the diffusion rates of the most important atomic
displacements can be predicted and used as input in KMC
simulations which are currently in progress.

\acknowledgments

It is our pleasure to thank L. Douillard and L. Proville for
stimulating discussions.

\appendix

\section{}

\begin{figure}[!fht]
\begin{center}
\includegraphics*[scale=0.6,angle=0]{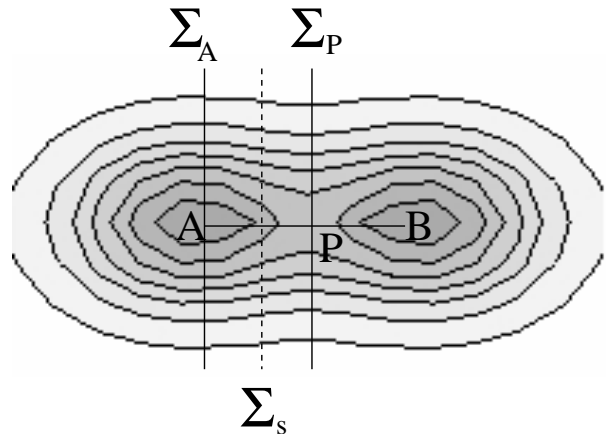}
\end{center}
\caption{Schematic potential energy surfaces in the vicinity of
two potential wells A and B separated by a saddle point P. The
surfaces $\Sigma_s$ are given by s=cst where s is the coordinate
along the diffusion path.} \label{fig:potential_energy_surface}
\end{figure}

Let us consider the (3N+3-dimensional) configuration space of a
system with N+1 identical atoms of mass $m$ and assume that the
potential energy of this system has two minima at points A and B
separated by a saddle point P. We suppose that the dividing
surface $\Sigma_P$ (containing P) between the regions A and B as
well as the diffusion path (i.e., the steepest descent line going
from A to B through P) have been determined (see Fig.\ref{fig:potential_energy_surface}). 
The coordinate system
is chosen so that one coordinate, $s$, runs along the diffusion
path. The equation of $\Sigma_P$ is then $s=s_P$. In the
transition state theory (TST) the diffusion rate from A to B is
given by \cite{Wahnstrom90}:

\begin{equation}
\Gamma_{TST}^{A\rightarrow B}(T)=\frac{<v_s
\theta(v_s)\delta(s-s_P)>}{<\theta(s_P-s)>} \label{eq:GTSTAB}
\end{equation}

\noindent where $v_s$ is the velocity along $s$, $<...>$ denotes a canonical average
and $\theta$ is the Heaviside function.

The basic assumption leading to Eq.\ref{eq:GTSTAB} is that each
crossing of $\Sigma_P$ corresponds to a diffusion event. This is
not true since recrossing may occur before thermalization in
region B. To correct for this effect a transmission coefficient
$\kappa(T)$ is introduced so that the total diffusion rate is:

\begin{equation}
\Gamma(T)=n_c \kappa(T) \Gamma_{TST}^{A\rightarrow B}(T)
\label{eq:GTT}
\end{equation}

\noindent when there are $n_c$ diffusion channels.

\subsection{Diffusion rate in the TST-HA}

When integrating over $p_s=mv_s$ in the canonical average of the
numerator of Eq.\ref{eq:GTSTAB}, $\Gamma_{TST}^{A\rightarrow
B}(T)$ can be rewritten:

\begin{equation}
\Gamma_{TST}^{A\rightarrow B}(T)=\frac{k_BT}{h}
\frac{Z_{\Sigma_P}}{Z_A} \label{eq:GTSTABZ}
\end{equation}

\noindent where $Z_{\Sigma_P}$ is a {\it constrained} partition
function, i.e., in which the representative points of the system
in the 3N+3-dimensional configuration space are compelled to stay
on $\Sigma_P$ and $Z_A$ is the partition function when the
representative points are located in the region around A. Thus the
dimensionality of the configuration space is 3N+2 for $Z_{\Sigma_P}$
and 3N+3 for $Z_A$.

If the static barrier $\Delta E$ is much larger than $k_BT$, the HA can be used
and $\Gamma_{TST}^{A\rightarrow B}(T)$ becomes:

\begin{equation}
\Gamma_{TST}^{A\rightarrow B}(T)=\frac{k_BT}{h}\exp(-(\Delta E + \Delta F_{vib})/k_BT)
\label{eq:GTSTABHA}
\end{equation}

\noindent where $\Delta F_{vib}=\Delta U_{vib}-T\Delta S_{vib}$ is
the contribution of vibrations to the free energy, with:

\begin{eqnarray}
\Delta U_{vib} & = &
\int_0^{\nu_{max}}\frac{h\nu}{2} \coth \big(\frac{h\nu}{2k_BT}\big)\Delta
n(\nu)d\nu \nonumber \\
\Delta S_{vib} & = &
k_B\int_0^{\nu_{max}}\bigg[ \frac{h\nu}{2k_BT} \coth \big ( \frac{h\nu}{2k_BT}\big)
-\ln \big(2 \sinh (\frac{h\nu}{2k_BT}) \big) \bigg]\Delta n(\nu)d\nu \label{eq:US}
\end{eqnarray}

\noindent and:

\begin{equation}
\int_0^{\nu_{max}}\Delta n(\nu)d\nu=-1.
\end{equation}

Indeed, in the HA the constraint on $Z_{\Sigma_P}$ can be
expressed as $q_P=0$ where $q_P$ is the normal coordinate
corresponding to the unstable mode at point P. As a consequence
this mode must be excluded from the summation over $\nu$.
Consequently, in this approximation and in the classical limit:

\begin{eqnarray}
\Delta U_{vib} & = & -k_BT  \\
\Delta S_{vib} & = & -k_B(1-\ln\frac{h\nu_0}{k_BT})
\label{eq:UScla}
\end{eqnarray}

\noindent where $\nu_0$ is the Vineyard attempt frequency
(Eq.\ref{nu0}). These quantities have been calculated explicitly
for self-diffusion on the (100) surface of Cu and Ag
\cite{Kurpick97b,Kurpick98}. Finally note that the correction
factor $\kappa(T)$ is generally omitted in the HA.

\subsection{Diffusion rate from the TI method}

In this approach, the integration over all momenta in both canonical averages
of Eq.\ref{eq:GTSTAB} is carried out in the classical limit. This leads to

\begin{equation}
 \Gamma_{TST}^{A\rightarrow B}(T)=(\frac{k_BT}{2\pi m})^{1/2}
\frac{\displaystyle \int_{\Sigma_P}\exp(-E(...r_{ij}...)/k_BT) d\Sigma_P}
{\displaystyle \int_{V_A}\exp(-E(...r_{ij}...)/k_BT) dV_A}. \label{eq:GTSTABTI}
\end{equation}

\noindent The integration domain extends over the hyper-surface
$\Sigma_P$ in the numerator and the hyper-volume $V_A$ limited by
$\Sigma_P$ and containing A in the denominator. Let $\Sigma_s$ be
the surface $s=cst$, i.e., it is normal at $s$ to the steepest
descent line containing A and P and set:

\begin{equation}
C\exp(-W(s)/k_BT)=\int_{\Sigma_s}\exp(-E(...r_{ij}...)/k_BT)
d\Sigma_s \label{eq:W}
\end{equation}

\noindent where C is a constant. If we multiply the numerator and denominator
of Eq.\ref{eq:GTSTABTI} by $\exp(-W(s_A)/k_BT)$, this equation can be rewritten as:

\begin{eqnarray}
\Gamma_{TST}^{A\rightarrow B}(T) &=& (\frac{k_BT}{2\pi
m})^{1/2}\Big[\int_{-\infty}^{s_P}\exp(-(W(s)-W(s_A))/k_BT)ds\Big]^{-1}
\exp(-(W(s_P)-W(s_A))/k_BT) \\  &=& \nu(T)\exp(-\Delta W/k_BT)
\label{eq:GBoisvert}
\end{eqnarray}

The constant C can be fixed by setting $W(s_A)=0$. Thus

\begin{equation}
W(s)=-k_BT\ln\frac{\displaystyle \int_{\Sigma_s}\exp(-E(...r_{ij}...)/k_BT)d\Sigma_s}
                  {\displaystyle \int_{\Sigma_A}\exp(-E(...r_{ij}...)/k_BT) d\Sigma_A}
\label{eq:Wnorm}
\end{equation}

The function $W(s)$ is known as the ``potential of mean force'' \cite{Roux91} in the
literature. Indeed it is straightforward to show that

\begin{equation}
\frac{dW}{ds}= <\frac{\partial E}{\partial s}>.
\label{eq:meanforce}
\end{equation}

Finally $\Delta W=W(s_P)-W(s_A)$ is an activation free energy
which satisfies:

\begin{equation}
\exp(-\Delta W/k_BT)=\frac{Z_{\Sigma_P}}{Z_{\Sigma_A}}
\label{eq:ZSZA}
\end{equation}

\noindent in which both $Z_{\Sigma_P}$ and $Z_{\Sigma_A}$ are now
{\it constrained} partition functions having thus the same
dimensionality (3N+2) of the configuration space. Note that this
point of view is closely related to the formulation of Wert and
Zener \cite{Wert49,Vineyard57}. We must emphasize that if we set
$\Delta W=\Delta E + \Delta U^{TI}- T\Delta S^{TI}$, the
quantities $\Delta U^{TI}$ and $\Delta S^{TI}$ are different from
$\Delta U_{vib}$ and $\Delta S_{vib}$ (Eq.\ref{eq:UScla}) even in
the HA. Indeed the number of modes, being the same in the two
partition functions, $\Delta U^{TI}_{HA}$ vanishes and $\Delta
S^{TI}_{HA}$ is independent of temperature.

This method has been used by Boisvert {\it et
al.}\cite{Boisvert98} for the diffusion of Cu on Cu(100) without
resorting to the HA. They used MD simulations to calculate $\Delta
W, \nu(T)$ and $\kappa(T)$ and showed that their result for
$\Gamma(T)$ (Eq.\ref{eq:GTT}) can be fitted nicely by an Arrhenius
law. Indeed they found that: {\it i)} if $\Delta W$ is written as
$\Delta E^{TI} - T\Delta S^{TI}$, then $\Delta E^{TI}$ and $\Delta
S^{TI}$ are both {\it effectively} temperature independent, {\it
ii)} $\kappa(T)$ and $\nu(T)$ are only slightly dependent on
temperature.

\end{document}